# Magnetoresistance of 2D and 3D Electron Gas in LaAlO$_3$/SrTiO$_3$ Heterostructures: Influence of Magnetic Ordering, Interface Scattering and Dimensionality

X. Wang[1,2], W.M Lü[1,2], A. Annadi[1,2], Z.Q. Liu[1,2], S. Dhar[1,3], T. Venkatesan[1,2,3], Ariando[1,2]

[1]NUSNNI-NanoCore, National University of Singapore, 117411 Singapore

[2]Department of Physics, National University of Singapore, 117542 Singapore

[3]Department of Electrical and Computer Engineering, National University of Singapore, 117576 Singapore

**Magnetoresistance (MR) anisotropy in LaAlO$_3$/SrTiO$_3$ (LAO/STO) interfaces is compared between samples prepared in high oxygen partial pressure ($P_{O2}$) of $10^{-4}$ mbar exhibiting quasi-two-dimensional (quasi-2D) electron gas and low $P_{O2}$ of $10^{-6}$ mbar exhibiting 3D conductivity. While MR of an order of magnitude larger was observed in low $P_{O2}$ samples compared to those of high $P_{O2}$ samples, large MR anisotropies were observed in both cases. The MR with the out-of-plane field is always larger compared to the MR with in-plane field suggesting lower dissipation of electrons from interface versus defect scattering. The quasi-2D interfaces show a negative MR at low temperatures while the 3D interfaces show positive MR for all temperatures. Furthermore, the angle relationship of MR anisotropy for these two different cases and temperature dependence of in-plane MR are also presented. Our study demonstrates that MR can be used to distinguish the dimensionality of the charge transport and various (defect, magnetic center, and interface boundary) scattering processes in this system.**

With the recent observation of electronic phase separation, strong negative MR and room temperature ferromagnetism at the LAO/STO interface prepared at higher $P_{O2}$ (~1×10$^{-2}$ mbar) [1], understanding the role of magnetic ordering and various scattering processes in the transport of quasi-2D electron gas has become important in addition to the variety of fascinating transport properties observed such as superconductivity, Kondo and field effect [2-6]. To differentiate between various magnetic states, the behavior of the MR can be studied as a function of field and the angle between the current and field direction [7-9]. For example, MR anisotropy measurements have been used to distinguish between weak localization and Kondo scattering [10]. In the case of the LAO/STO interfaces, several MR studies [11-14] have been done to study interface anisotropy, long-range magnetism, magnetic inhomogeneities and spin-orbit scattering. Based on the observation of negative MR, Shalom *et al.* [14] proposed the existence of magnetic ordering below 35 K for the samples prepared at the canonical $P_{O2}$ from 5x10$^{-5}$ to 1x10$^{-3}$ Torr where quasi-2D electron gas is dominant. However, there has not been any study of the comparison of MR between the quasi-2D and the 3D conductivity cases or the angular dependence of the MR as a function of field and the angle between the current and field direction. Such MR anisotropy measurements would be able to give us further information about magnetic ordering, transport dimensionality and various (interface) scattering processes that can exist in electronic transport of a confined system.

In this paper, we report on temperature dependent MR anisotropy measurement for atomically sharp interfaces between LaAlO$_3$ and SrTiO$_3$ not only for the quasi-2D electron gas prepared under the conditions of $P_{O2}$ = 10$^{-4}$ mbar, but also the 3D electron gas at $P_{O2}$ = 10$^{-6}$ mbar. Fundamental differences were observed in MR behavior of high $P_{O2}$ samples, where a quasi-2D electron gas is expected, compared to those of low $P_{O2}$ samples, where the electronic transport is 3D. In samples prepared at high $P_{O2}$, the MR behavior is strongly influenced by the existence of magnetic scattering centers near the interfaces (a magnetic scattering plane), too small to be detected by other means which accounts for a negative low temperature MR, and the scattering most probably due to the vicinal steps of the substrate which accounts for the linear MR observed even for out-of-plane configuration.

In our experiment, samples of 26 unit cells (uc) $LaAlO_3$ were grown layer-by-layer on atomically flat $TiO_2$ terminated $SrTiO_3$, under two different $P_{O2}$ of $10^{-4}$ and $10^{-6}$ mbar. Reflection high energy electron diffraction (RHEED) oscillations obtained from both types of interface samples during growth are shown in Fig. 1a and 1b. These oscillations indicate good layer-by-layer growth up to the 26 uc thick of $LaAlO_3$. The electrical measurement was done by linear four-probe geometry.

Figure 1c shows the resistances of the two types of interfaces. A large difference of 2 orders of magnitude in resistance values between the 2D and 3D samples was observed at room temperature and increases to 4 orders at 2 K. As commonly believed, the resistance difference is one of the key differences between the quasi-2D and 3D interfaces with the carriers in the latter case arising from oxygen vacancies. Furthermore the growth pressure of $10^{-4}$ mbar also matches well with reported critical $P_{O2}$, above which $SrTiO_3$, when annealed, will remain an insulator and conductance is generally from interface [15].

MR anisotropy investigation was done in linear geometry with two different directions of magnetic field applied, namely, in-plane MR and out-of-plane MR as shown in Fig. 2a and 2b. One thing to be noticed is that the magnetic fields were always applied perpendicular to current. The field was in the plane of the film at 0 degrees and normal to the film at 90 degrees.

For the oxygen vacancies dominated 3D interfaces, a very large out-of-plane MR of ~1500% (Fig. 2c) and an order of magnitude lower in-plane MR of ~100% (Fig. 2d) were observed when magnetic field was increased up to 9 Tesla (T) at 2 K. While the out-of-plane MR has a quadratic relation, the in-plane MR has a linear relation with applied magnetic field. The behavior of the out-of-plane MR (quadratic dependence) is understood as due to increased defect scattering resulting from enhanced transit path of electrons [16]. On the other hand the in-plane MR is mainly dominated by interface scattering, primarily at the LAO/STO interface as the estimated cyclotron radius for the electron is of the order of macrometer at a 1 T field which is significantly larger than the thickness of the $LaAlO_3$ layer. As a result the frequency of the interface scattering will be proportional to the cyclotron frequency which has a linear dependence on magnetic field. Further, the reduced magnitude of the MR indicates that the interface scattering is significantly less dissipative (elastic scattering) than the defect scattering.

The MR at 2 K for the quasi-2D interfaces shows much more interesting phenomena. The out-of-plane MR in quasi-2D interfaces is linear instead of quadratic (Fig. 2e). The magnitude of the MR is also an order of magnitude smaller compared to the 3D case under 9 T and closer to the case of in-plane 3D MR. Both observations support the idea of the 2D electrons scattering from the vicinal steps in $SrTiO_3$ which have a width of the order of 200 nm, significantly smaller than the cyclotron radius. Surprisingly, a negative MR (Fig. 2f) is observed for the in-plane geometry, in contrast to all the other cases. The negative MR could be an indication of onset of magnetic centers as the scattering becomes more coherent. The origin of magnetic scattering has been seen before in the form of Kondo scattering and also in the extreme case of interfaces grown under higher oxygen pressures exhibiting electronic phase separation [1]. The origin of the magnetic centers is most likely from cationic defects at the LAO/STO interface in the form of Ti vacancies or $Ti^{3+}$. Nakagawa and Hwang *et al*. have used electron energy loss spectroscopy measurement of the interface to show the existence of $Ti^{3+}$ [17]. Thus one expects a quasi-2D plane of magnetic centers near the interface responsible for the negative MR.

To sum up, there are three kinds of MR relations observed at 2 K: quadratic positive MR arising from enhanced electron transit path (out-of-plane MR in low $P_{O2}$ sample), linear MR arising from interface scattering (in-plane MR in low $P_{O2}$ sample and out-of-plane MR in high $P_{O2}$ sample), and negative MR arising from coherent scattering (in-plane MR in high $P_{O2}$ sample).

The anisotropy features could have been seen from previous Fig. 2. In order to check the detailed features of those anisotropies, the MR under 9T for both the quasi-2D and 3D interfaces are measured at different angles and different temperatures and are shown in Fig. 3. The MR and the anisotropy for both quasi-2D and 3D interfaces are suppressed at temperatures above ~100 K. At lower temperatures for the 3D interfaces, the resistance exhibits a rough cosine relationship with respect to measured angle and the out-of-plane (0 degree) resistance is about 4 times larger than in-plane (90 degree) resistance at 9 T and 2 K. In the quasi-2D interface the functional form of the angular dependence shows the formation of a deep cusp at 90 and 270 degrees which is characteristic of a 2D electron transport [18].

Figure 4 shows a closer comparison of the angular dependence of the 2D and the 3D cases at 2 K and 9 T field. In the quasi-2D interfaces, a negative MR is observed in in-plane geometry (90 degree) and positive MR in out-of-plane geometry (0 degree). To vividly demonstrate the MR anisotropy of different types of interfaces, two different plots were used to present the MR anisotropy at 2 K under 9 T. As can be seen in Fig. 4, the negative MR in-plane geometry for quasi-2D interfaces is very obvious. The shape differences and amplitude differences between quasi-2D and 3D interfaces could be clearly observed in these plots.

The observed different scattering mechanisms have also temperature dependences as illustrated in in-plane resistance and in-plane MR shown in Fig. 5 for the 2D interface. For the 2D samples prepared at higher pressures of $10^{-3}$ mbar Kondo effect has been clearly seen but not at $10^{-4}$ mbar as the concentration of the magnetic centers is too low. However, a magnetic field can align these residual centers, which induces a more coherent scattering and results in a negative MR. Thus the negative MR is an even more sensitive probe for the presence of magnetic centers than Kondo scattering. This negative MR vanishes beyond 20 K turning progressively positive at higher temperatures due to the disruption of the exchange interaction between the magnetic centers by thermal excitations. This accounts for the downward trend of the MR signal with decreasing temperature originates at 30 K and becomes negative below 20

K. However, the abruptness of the MR transition at 30 K tends to suggest the role of the SrTiO$_3$ phase transition (orthorhombic to rhombohedral) on the observed change in the MR behavior. Further study will be required to elucidate this.

In summary, we present a comparison of MR anisotropy in LAO/STO interfaces prepared under different $P_{O2}$. Large anisotropies were found in both 2D and 3D interface samples and three distinct scattering mechanisms were observed in this system. The observed anisotropy features and temperature dependence suggest the role of interface scattering in addition to enhanced electron paths under a magnetic field and in the case of 2D electron gas the role of a magnetic plane that is effective at low temperatures in introducing a coherent scattering process leading to negative MR. This study supports the formation of a magnetic scattering plane near the 2D electron interface. The study also brings out the lower dissipation of scattering at interfaces as opposed to defect scattering losses due to enhanced electron transit paths. Thus MR anisotropy is a sensitive technique for understanding the role of magnetic ordering and various scattering processes in the transport of quasi-2D electron gas.

**Acknowledgement**

We thank J. Huijben and H. Hilgenkamp for experimental help and discussions and the National Research Foundation (NRF) Singapore under the Competitive Research Program 'Tailoring Oxide Electronics by Atomic Control', NUS cross-faculty grant and FRC for financial support.

**Figure Captions:**

**Figure 1:** RHEED oscillation for samples prepared in (a) high $P_{O2}$ and (b) low $P_{O2}$. Clear layer-by-layer growth was observed in both cases. (c) Large transport resistance difference for samples processed under different $P_{O2}$.

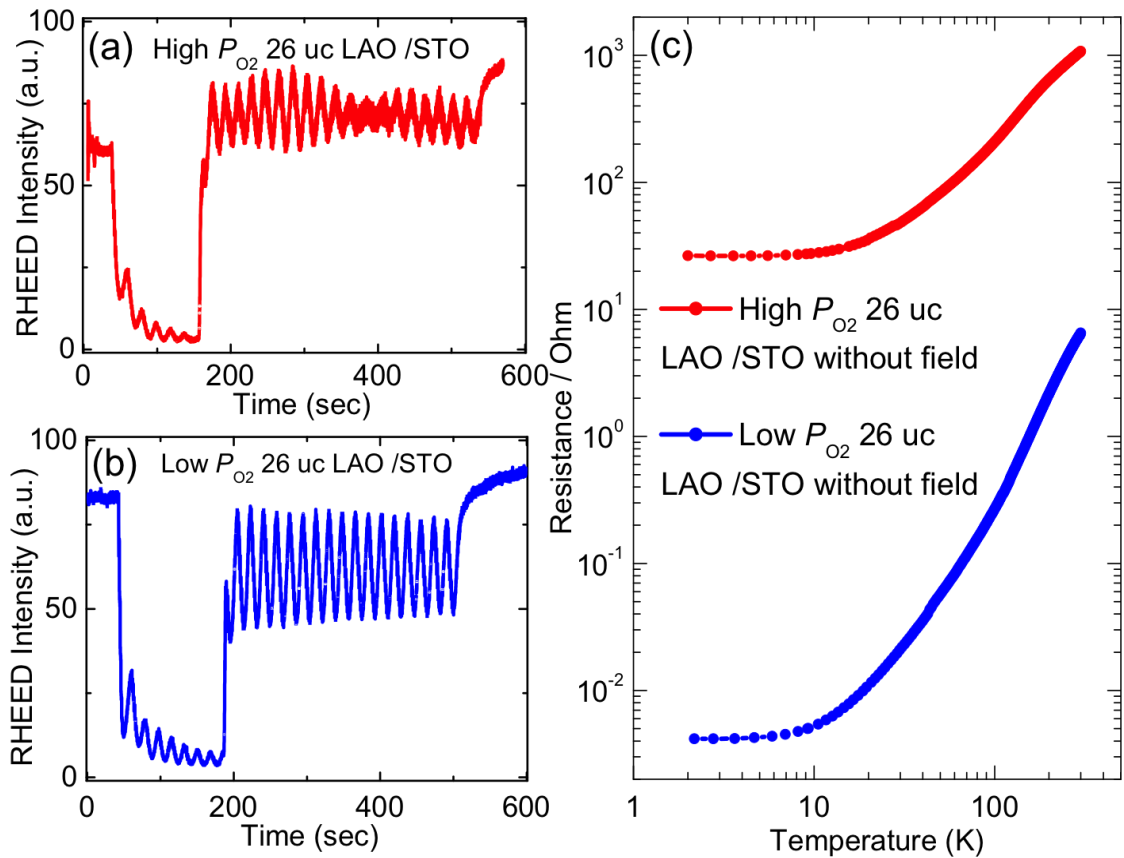

**Figure 2:** Comparison on magnetoresistance between high $P_{O2}$ and low $P_{O2}$ with magnetic field applied at different directions at 2K. Illustrations for (a) out-of-plane and (b) in-plane linear measurement geometry and MR for four cases (c) Low $P_{O2}$ LAO/STO interfaces out-of-plane MR, (d) Low $P_{O2}$ LAO/STO interface in-plane MR, (e) High $P_{O2}$ interfaces out-of-plane MR, and (f) High $P_{O2}$ interfaces in-plane MR.

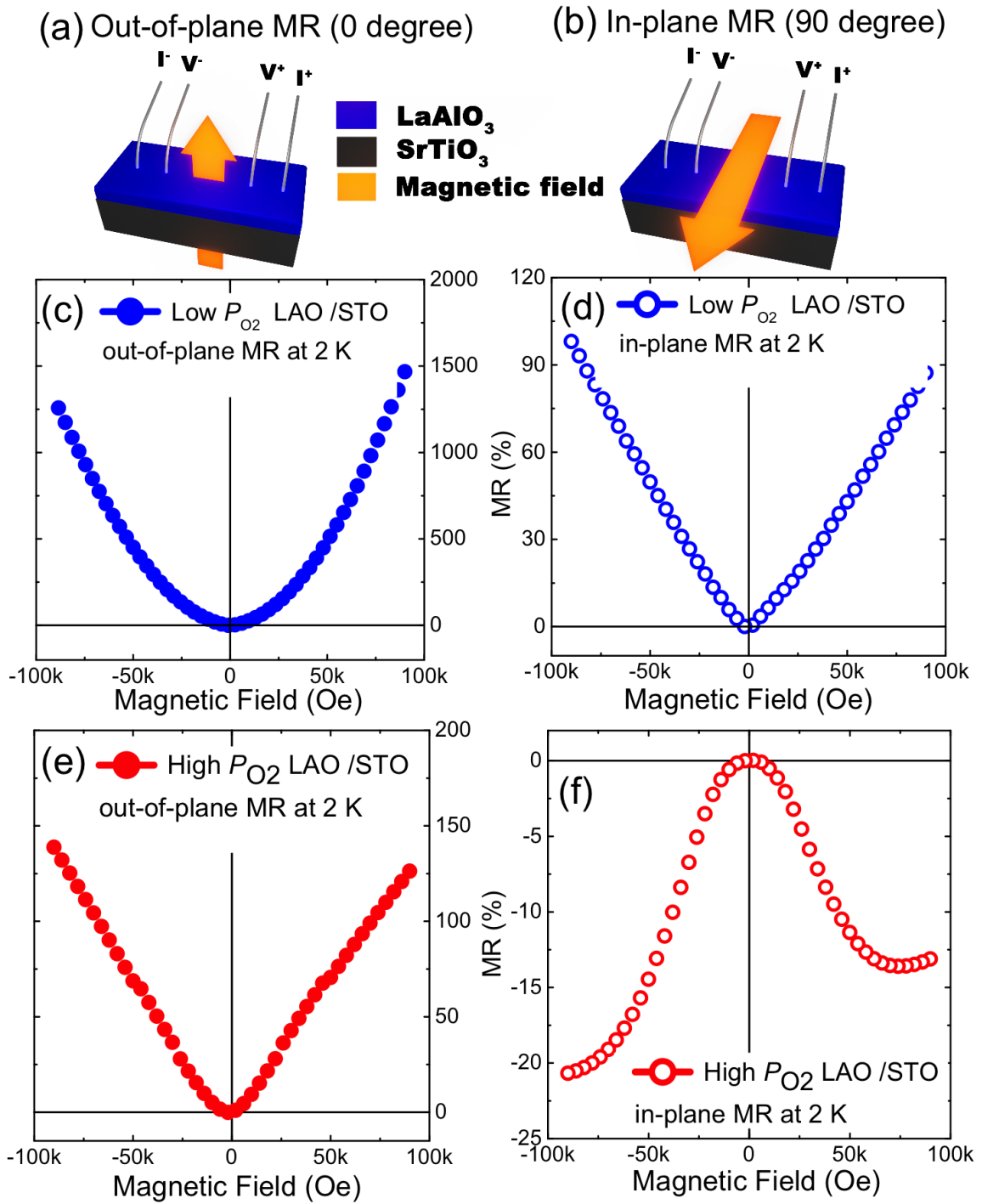

**Figure 3:** Resistance under 9 T magnetic field with respect to different angle for two types of interfaces.

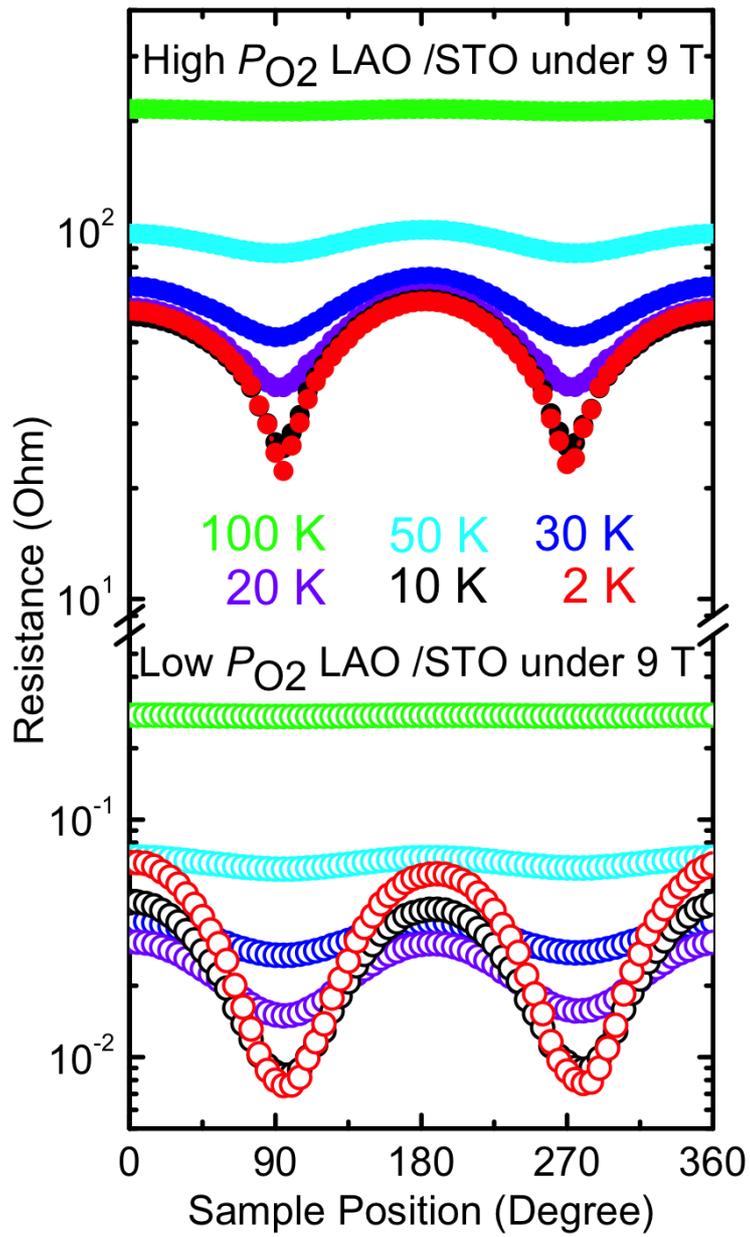

**Figure 4:** Various plots for magnetoresistance of different interfaces under 9 T magnetic field at 2 K. Normal plot for quasi-2D interfaces (a) and 3D interfaces (b); Polar plot for quasi-2D interfaces (c) and 3D interfaces (d).

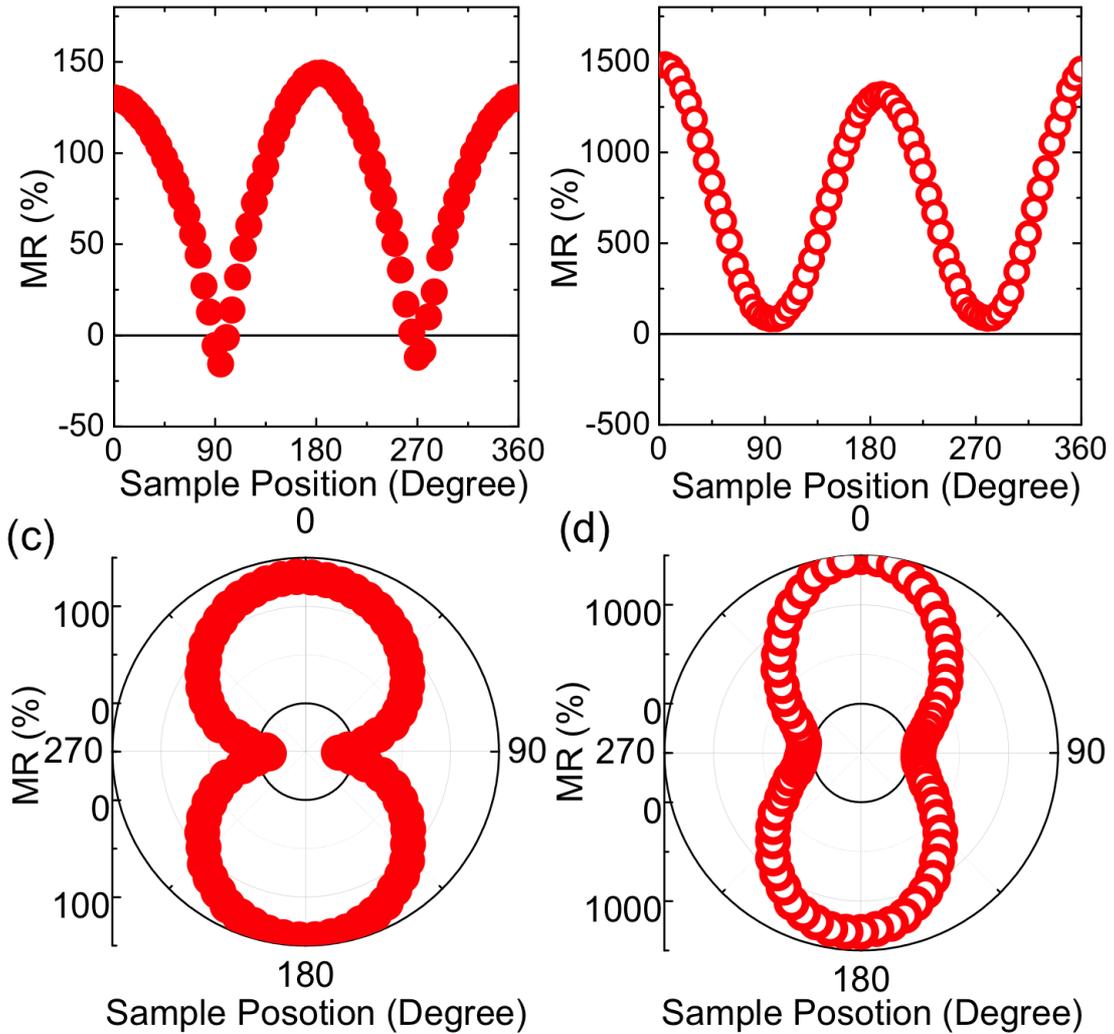

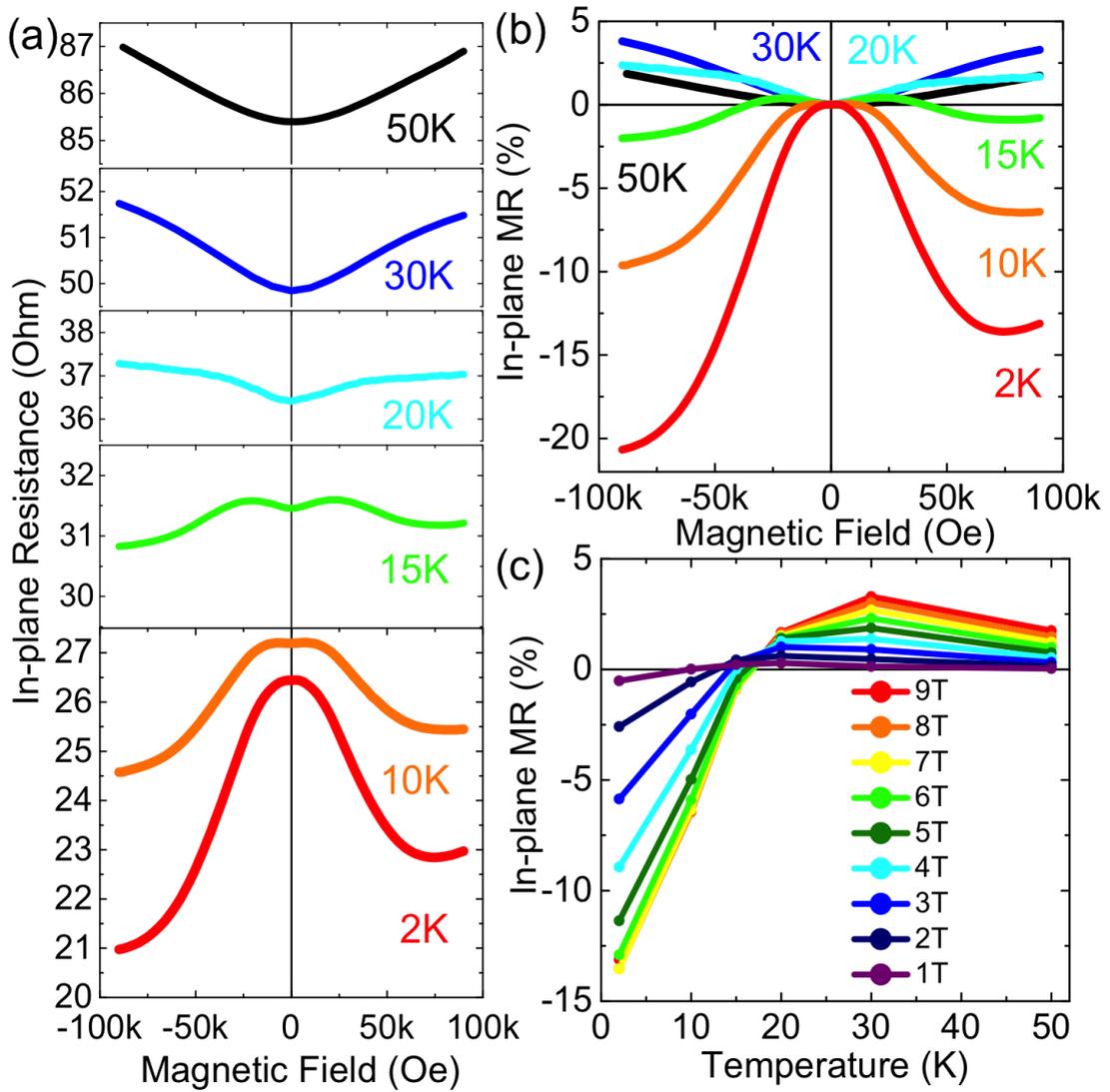

**Figure 5:** Temperature dependence for in-plane resistance (a) and in-plane magnetoresistance (b and c) for high $P_{O2}$ LAO/STO interfaces.